\documentclass[twoside]{article}
\usepackage{fleqn,espcrc2,latexsym,bbm,feynmp,graphicx,epsfig}

\newcommand{\partialslash}{\partial\!\!\!\!\!\!\not\,\,}
\newcommand{\pslash}{p\!\!\!\!\!\not\,\,}

\newcommand{\Gslash}{G\!\!\!\!\!\!\not\,\,}
\newcommand{\Gtildeslash}{\tilde{G}\!\!\!\!\!\!\not\,\,}
\newcommand{\Aslash}{A\!\!\!\!\!\!\not\,\,}

\newcommand{\AmS}{{\protect\the\textfont2
  A\kern-.1667em\lower.5ex\hbox{M}\kern-.125emS}}

\title{Inclusive Quark Production in $e^{+}e^{-}$-Annihilation - \\A Path Integral Approach}

\author{O. Nachtmann, A. Rauscher\address{Institut f\"ur Theoretische Physik der Universit\"at Heidelberg, Philosophenweg 16, D-69120 Heidelberg}\thanks{supported by the German Bundesministerium f\"ur Bildung und Forschung (BMBF), Contract no. 05 7HD 91 P(0), by Studienstiftung des Deutschen Volkes and by Deutsche Forschungsgemeinschaft (DFG), Grant no. GKR 216/1-9}}
       
\begin{document}

\begin{abstract}
The single-particle inclusive differential cross-section for a reaction $a+b\to c+X$ is written as imaginary part of a correlation function in a forward scattering amplitude for $a+b\to a+b$ in a modified effective theory. In this modified theory the interaction Hamiltonian $\tilde H_I$ equals $H_I$ of the original theory up to a certain time. Then there is a sign change and $\tilde H_I$ becomes nonlocal. This is worked out in detail for scalar field models and for QED plus the abelian gluon model. A suitable path integral for direct calculations of inclusive cross sections is presented.
\vspace{1pc}
\end{abstract}

\maketitle

\begin{fmffile}{mpbild}

\section{INTRODUCTION}

Inclusive cross sections like
\begin{eqnarray}
e^{+}+e^{-}\to\pi^\pm+X\nonumber
\end{eqnarray}
are measured at a centre-of-mass-energy $\sqrt{s}$=91.2 GeV for example by the DELPHI detector at LEP\cite{0}. So far, only phenomenological models, which need to be tuned to the data, are available to describe the fragmentation of quarks and gluons into hadrons. A direct QCD calculation for these inclusive cross sections seems difficult because of the importance of non-perturbative effects. In order to be able to include these soft processes, one should start from an expression for the inclusive cross section which is suitable for non-perturbative calculations.

\medskip

The article is organised as follows. In section 2 the basic relations for single inclusive cross sections are recalled. In section 3 we present our general formalism for the modified effective theory in the case of scalar fields. As a specific example the cross section $e^{+}+e^{-}\to q+X$ is considered in quantum electrodynamics (QED) with massive photons and the theory of quarks interacting with abelian gluons in section 4. A path integral representation for inclusive cross sections in the abelian gluon model is derived. 

\section{SINGLE INCLUSIVE CROSS SEC-\\TIONS}

In this section we recall some basic relations for inclusive cross sections. Our notation follows \cite{4,4a}. Let us consider a single-particle inclusive reaction, i.e
\begin{equation}\label{2.1}
a(p_{1})+b(p_{2})\to c(p_{3})+ X(p_{X}).
\end{equation}
To take a simple case, let $a,b,c$ be spinless particles with masses $m_a, m_b, m_c$. The c.m. energy squared is $s=(p_1+p_2)^2$. The $S$-matrix element for the reaction (\ref{2.1}) is given as
\begin{eqnarray}\label{2.2}
S_{fi}
&=&\langle\ c(p_{3}), X(p_{X}), out\ |\ a(p_{1}), b(p_{2}), in\ \rangle\nonumber\\
&=&iZ_{c}^{-1/2}\int d^4\!x\ e^{ixp_{3}} \left(\Box_{x}+m^2_c\right)\nonumber\\
&&\langle\ X(p_{X}), out\ |\ \phi_{c}(x)\ |\ a(p_{1}), b(p_{2}), in\ \rangle.\nonumber\\
\end{eqnarray}
Here we have applied the reduction formula for particle $c$ in the final state. Let $\phi_c(x)$ be a suitable interpolating field for $c$ and $Z_c$ the corresponding wave function renormalisation constant. The ${\cal T}$-matrix element is obtained from the $S$-matrix element via
\begin{eqnarray}
S_{fi}&=&\delta_{fi}+i(2\pi)^4\delta^{(4)}(p_{1}+p_{2}-p_{3}-p_{X}){\cal T}_{fi}\nonumber
\end{eqnarray}
\begin{eqnarray}
{\cal T}_{fi}&=&Z_c^{-1/2}\left(\Box_{y}+m^2_c\right)\langle\ X(p_X), out\ |\ \nonumber\\
&&\qquad\phi_c(y)\ |\ a(p_1), b(p_2), in\ \rangle\raisebox{-1ex}{\em $\mid_{y\to0^{-}}$}.
\end{eqnarray}
The single-particle inclusive cross section $f_{inc}(p_{3})$ is defined by
\begin{eqnarray}\label{2.5}
f_{inc}(p_{3})&:=&p_{3}^0\frac{d^3\sigma}{d^3p_3}(a+b\to c+X)\nonumber\\
&\ =&\frac{1}{4(2\pi)^3w(s,m^2_a,m^2_b)}\sum_X\nonumber\\&&(2\pi)^4\delta^{(4)}(p_1+p_2-p_3-p_X)|{\cal T}_{fi}|^2.\nonumber\\
\end{eqnarray}
In the usual way the sum over all states $|X, out\ \rangle$ in (\ref{2.5}) can be carried out using completeness and translational invariance
\begin{eqnarray}
f_{inc}(p_{3})&=&\frac{1}{4(2\pi)^3w}Z^{-1}_c\int d^4\!x\ e^{ip_3x}\nonumber\\
&&\left(\Box_{y}+m^2_c\right)\left(\Box_{z}+m^2_c\right)\nonumber\\
&&\langle a(p_1),b(p_2), in\ |\phi_c^\dagger(y)\phi_c(x+z)\nonumber\\
&&\qquad|a(p_1),b(p_2), in\ \rangle\raisebox{-1ex}{\em $\mid_{y\to0^{-},\ z\to0^{-}}$}.\nonumber\\
\end{eqnarray}
Rewrite $f_{inc}(p_{3})$ to get time-ordered operators
\begin{eqnarray}
f_{inc}(p_{3})=\frac{1}{2(2\pi)^3w}\ \mbox{Im}\ {\cal C}(p_1,p_2,p_3),
\end{eqnarray}
\begin{eqnarray}
{\cal C}(p_1,p_2,p_3)=i\int d^4\!x\ e^{ip_3x}{\cal M}(x)\label{2.8},
\end{eqnarray}
\begin{eqnarray}
{\cal M}(x)&=&Z^{-1}_c\left(\Box_{y}+m^2_c\right)\left(\Box_{z}+m^2_c\right)\theta(-x^0)\nonumber\\
&&\langle a(p_1),b(p_2), in\ |\ \phi_c^\dagger(y)\phi_c(x+z)\nonumber\\
&&\qquad|a(p_1),b(p_2), in\ \rangle\raisebox{-1ex}{\em $\mid_{y\to0^{-},\ z\to0^{-}}$}.\nonumber\\
\end{eqnarray}
The amplitude ${\cal C}(p_1,p_2,p_3)$ will be written as a field-field correlation function in a forward scattering amplitude, but in a modified effective theory.

\section{MODIFIED EFFECTIVE THEORY\\ FOR SCALAR FIELDS}

Let us assume that the basic dynamical variables of the original theory are the operators for unrenormalised scalar fields $\phi_i(x)$ and their conjugate canonical momenta $\Pi_i(x)\ (i=1,...,N)$. We denote $\phi_i(x),\Pi_i(x)$ collectively as $\Phi(x)$. Let $H$ be the Hamiltonian of the system which we split into a free part $H_0$ and an interaction part $H_I$ which may depend explicitly on the time $t$, but should not involve time derivatives of $\Pi_i(x)$
\begin{equation}\label{3.1}
H(t,\Phi(\vec{x},t))=H_{0}(\Phi(\vec{x},t))+H_{I}(t,\Phi(\vec{x},t))
\end{equation}
with
\begin{eqnarray}
\lim_{t\to\pm\infty}H(t,\Phi(\vec{x},t))={H_{0}}(\Phi(\vec{x},t)).
\end{eqnarray}
Besides the interacting fields and momenta $\Phi$ free fields and momenta $\Phi^{(0)}$ are considered with the corresponding Hamiltonian $H_0$.

\medskip

We assume now as usual that there exist uni\-tary operators $U(t)$ that realize the time-de\-pen\-dent canonical transformations relating $\Phi$ to $\Phi^{(0)}$
\begin{equation}\label{3.2}
\Phi(\vec{x},t)=U^{-1}(t)\Phi^{(0)}(\vec{x},t)U(t).
\end{equation}
Taking as boundary condition
\begin{equation}\label{3.2a}
\lim_{t\to-\infty}\Phi(\vec x,t)=\Phi^{(0)}(\vec x,t)
\end{equation}
we get
\begin{eqnarray}\label{eqn111}
U(t)=\mbox{T}\exp\left[-i\int_{-\infty}^{t}dt'H_{I}(t',\Phi^{(0)}(\vec{x},t'))\right].
\end{eqnarray}
We define furthermore
\begin{eqnarray}\label{3.5}
U(t_{2},t_{1})=U(t_{2})U^{-1}(t_{1}).
\end{eqnarray}
so that the $S$-matrix is given as
\begin{eqnarray}
S&=&\lim_{t\to-\infty}U(t,-t)\nonumber\\
&=&\mbox{T}\exp\left[-i\int_{-\infty}^{+\infty}dt'H_{I}(t',{\Phi^{(0)}(\vec{x},t')})\right].\nonumber\\
\end{eqnarray}

\medskip

Now we return to the single inclusive cross section, where we have to calculate the matrix element 
\begin{eqnarray}
{\cal M}(x)&=&Z^{-1}_c\left(\Box_{y}+m^2_c\right)\left(\Box_{z}+m^2_c\right)\theta(-x^0)\nonumber\\
&&{\langle a(p_1),b(p_2), in\ |}\ \phi_c^\dagger(y)\phi_c(x+z)\nonumber\\
&&\qquad|a(p_1),b(p_2), in\ \rangle\raisebox{-1ex}{\em $\mid_{y\to0^{-},\ z\to0^{-}}$}.\nonumber\\
\end{eqnarray}
Following the time-dependence in ${\cal M}(x)$ from the right to the left, we start at time $-T\to-\infty$ and pass through operators of increasing time arguments until time $0$. Then the time sequence changes and we go back in time to time $-T'\to-\infty$. In a usual matrix element the time arguments should increase instead. We will now show that we can write the matrix element ${\cal M}(x)$ in the usual form, with time increasing from right to left, if we pass to operators $\tilde{U}$ of the form (\ref{eqn111}) but with a modified interaction Hamiltonian $\tilde{H}_{I}$. As the time-sequence in ${\cal M}$ is correct up to $t=0$, we request
\begin{eqnarray}
{1)}&&{\tilde{H}_{I}(t)=H_{I}(t,\Phi^{(0)}(\vec{x},t))\quad\mbox{for}\quad t<0}\nonumber\\
{2)}&&{\tilde{S}={\mathbbm 1}}
\end{eqnarray}
with 
\begin{eqnarray}
\tilde{S}=\mbox{T}\exp\left[-i\int_{-\infty}^{+\infty}dt'\tilde{H}_{I}(t',{\Phi^{(0)}(\vec{x},t')})\right].
\end{eqnarray}
It follows
\begin{eqnarray}
\tilde{H}_{I}(t)&=&\theta(-t)H_{I}(t,{\Phi^{(0)}(\vec{x},t)})\nonumber\\
&&\quad-\theta(t)H_{I}({-t},{\Phi^{(0)}(\vec{x},{-t})}).
\end{eqnarray}
For $t>0$ our modified interaction Hamiltonian $\tilde H_I(t)$ depends on the free fields and momenta at time $(-t)$. But we know how to express the free fields and momenta at time $(-t)$ by their values at time $t$ using the free field equations of motion. Thus, we can consider $\tilde H_I(t)$ as a nonlocal functional of the dynamical variables $\Phi^{(0)}(\vec x,t)$ at the same time $t$ also for $t>0$
\begin{eqnarray}\label{3.21}
\tilde H_I(t)=\tilde H_I(t,\Phi^{(0)}(\vec x,t)).
\end{eqnarray}

\medskip

Therefore, we can consider our matrix element as of the standard type but in the modified theory governed by the total Hamiltonian
\begin{eqnarray}\label{3.25}
\tilde H(t,\tilde\Phi(\vec x,t))=H_0(\tilde\Phi(\vec x,t))+\tilde
H_I(t,\tilde\Phi(\vec x,t)),
\end{eqnarray}
Here we denote by $\langle\!\langle\ \rangle\!\rangle$ matrix elements in the modified theory 
\begin{eqnarray}
{\cal M}(x)&=&Z_{c}^{-1}\theta(-x^0)\left(\Box_{y}+m^2_c\right)\left(\Box_{z}+m^2_c\right)\nonumber\\
&&{\langle\!\langle\ a(p_1), b(p_2), out\ |\ \phi_c^\dagger(y)\phi_c(x+z)\ }\nonumber\\
&&\qquad{|\ a(p_1), b(p_2), in\ \rangle\!\rangle} \raisebox{-1ex}{\em $\mid_{y\to0^{-},\ z\to0^{-}}$}.\nonumber\\
\end{eqnarray}

\medskip

The matrix element ${\cal M}$ is written in the standard form with an in-state to the right and an out-state to the left. The prize we have to pay is that we have to use the modified theory where the Hamiltonian $\tilde{H}(t)$ has a sudden variation at $t=0$ and is nonlocal for $t>0$. 

\section{INCLUSIVE PRODUCTION IN QED\\ AND AN ABELIAN GLUON MODEL}

\subsection{QED and the Abelian Gluon Model}

In this section inclusive reactions are considered in QED coupled to an abelian gluon model. An example for such a reaction is
\begin{eqnarray}
\label{eqn90}
e^{+}(p_{1})+e^{-}(p_{2})\ \to\ q(p_{3})+X.
\end{eqnarray}

\medskip

Starting point is the Lagrangian describing the interaction of electrons of mass $m$ and charge $-e$ with a massive photon of mass $\lambda$ - to avoid any infrared divergences - and of two quark flavours of equal mass $M$ and electric charge $eQ_{q}$ with the photon and a massive abelian gluon of mass $\mu$. As Lagrangian we choose
\begin{eqnarray}
\label{eqn1000}
{\cal L}&=&-\ \frac{1}{4}G_{\mu\nu}G^{\mu\nu}-\frac{1}{2\eta_{0}}(\partial_{\mu}G^{\mu})^2+\frac{1}{2}\mu_{0}^2G_{\mu}G^{\mu}\nonumber\\
&&-\ \frac{1}{4}F_{\mu\nu}F^{\mu\nu}-\frac{1}{2\xi_{0}}(\partial_{\mu}A^{\mu})^2+\frac{1}{2}\lambda_{0}^2A_{\mu}A^{\mu}\nonumber\\
&&+\ \bar{\psi}\left(\frac{i}{2}\!\!\!\!\stackrel{\ \ \leftrightarrow}{\partialslash}-m_{0}+e_{0}\Aslash\ \right)\psi\nonumber\\
&&+\bar{q}\left(\frac{i}{2}\!\!\!\!\stackrel{\ \ \leftrightarrow}{\partialslash}-M_{0}-e_{0}Q_{q}\Aslash-g_{0}\tau_{3}\Gslash\ \right)q,
\end{eqnarray}
where $G_{\mu}$ denotes the abelian gluon field and $G_{\mu\nu}=\partial_{\mu}G_{\nu}-\partial_{\nu}G_{\mu}$ its field strength tensor, $A_{\mu}$ the photon field and $F_{\mu\nu}=\partial_{\mu}A_{\nu}-\partial_{\nu}A_{\mu}$ its field strength tensor, $\psi$ the electron field and $q$ the quark field. We have a quark field with two flavours and $\tau_{3}$ is the usual Pauli matrix.

\medskip

The Hamiltonian corresponding to (\ref{eqn1000}) can be obtained as usual by a Legendre transformation. Only the interaction term has to be altered in order to construct the Hamilton operator for the modified theory 
\begin{eqnarray}\label{eqn16}
\tilde{H}_{I}(y^0)=\int\!d^3y\ [\theta(-y^0){\cal H}_{I}(y)-\theta(y^0)\tilde{\cal H}_{I}(\tilde{y})]\nonumber\\
\qquad\qquad\mbox{with}\ \tilde{y}=(-y^0,\vec{y}).\nonumber\\
\end{eqnarray}
In $\tilde{\cal H}_{I}(\tilde{y})$ the fields at time $-y^0$ have to be substituted by the fields at time $y^0$. Two points should be stressed: First, due to the jump in the interaction, the effective theory has no time translation invariance, so that no longer energy conservation holds at every vertex. Second, the effective interaction contains derivatives. 

\subsection{Inclusive Quark Production in $e^{+}e^{-}$-\\Annihilation}

For unpolarised $e^{-}$ and $e^{+}$ in the initial state and summation over spin and flavour of the final state quark the matrix element reads 
\begin{eqnarray}
{\cal M}(x)&=&-Z^{-1}_{q}\sum_{s_{q}}\bar{u}(p_{3})(i\overrightarrow{\partialslash_{z}}-M)\nonumber\\
&&\sum_{s_{e^{+}},s_{e^{-}}}\ \!\!\!\!\!\!'\!\langle\!\langle\ e^{+}(p_{1}),e^{-}(p_{2})\ |\ \mbox{T}q_{A}(x+z)\nonumber\\
&&\qquad\bar{q}_{A}(y)\ |\ e^{+}(p_{1}),e^{-}(p_{2}) \rangle\!\rangle\theta(-x^0)\nonumber\\
&&\qquad(-i\overleftarrow{\partialslash_{y}}-M)u(p_{3})\raisebox{-1ex}{\em $\mid_{y\to 0^{-},\ z\to 0^{-}}$}.\nonumber\\
\end{eqnarray}
In order to calculate ${\cal M}$ we expand it in powers of the electromagnetic coupling constant $e$. The electromagnetic interaction of the incoming fermions can be separated
\begin{eqnarray}
&&{\cal C}^{a}(p_{1},p_{2},p_{3})=Z^{-1}_{q}e^4Q_{q}^2l^{\mu\nu}\left\vert\frac{1}{s-\lambda^2+i\epsilon}\right\vert^2\nonumber\\
&&i\int d^4x e^{ip_{3}x}\theta(-x^0)\nonumber\\
&&\ \sum_{s_{q}}\bar{u}(p_{3})(i\overrightarrow{\partialslash_{z}}-M)\langle\!\langle 0 |\mbox{T}q_{A}(x+z)\nonumber\\
&&\qquad\int\!\! d^4\!x_{2}[\theta(-x_{2}^0)e^{-i(p_{1}+p_{2})x_{2}}\bar{q}(x_{2})\gamma_{\nu}q(x_{2})\nonumber\\
&&\qquad\qquad-\theta(x_{2}^0)e^{-i(p_{1}+p_{2})\tilde{x}_{2}}\bar{q}(\tilde{x}_{2})\gamma_{\nu}q(\tilde{x}_{2})]\nonumber\\
&&\qquad\int\!\! d^4\!x_{1}[\theta(-x_{1}^0)e^{i(p_{1}+p_{2})x_{1}}\bar{q}(x_{1})\gamma_{\mu}q(x_{1})\nonumber\\
&&\qquad\qquad-\theta(x_{1}^0)e^{i(p_{1}+p_{2})\tilde{x}_{1}}\bar{q}(\tilde{x}_{1})\gamma_{\mu}q(\tilde{x}_{1})]\nonumber\\
&&\quad\bar{q}_{A}(y) | 0 \rangle\!\rangle(-i\overleftarrow{\partialslash_{y}}-M)u(p_{3})\raisebox{-1ex}{\em $\mid_{y\to 0^{-},z\to 0^{-},e=0}$}\nonumber\\
\end{eqnarray}

As a simple check of our theoretical manipulations let us finally calculate the lowest order in the quark-gluon-coupling $g$, i.e. for $g=0$. Decomposing the quark 6-point-functions with Wick's theorem into quark 2-point-functions and substituting for these the perturbative propagators we get for ${\cal C}^{a}$ two contributions according to the two different possible contractions shown in figure 1.

\begin{figure}[ht]
\unitlength1mm
\begin{eqnarray}
&&\parbox{60mm}{\begin{fmfgraph*}(60,30)
\fmfstraight\fmftop{v5,v4,v2,v1,v3,v6}
\fmfleft{i1,i2}\fmfright{o1,o2}
\fmf{electron,label=$e^{-}$}{v8,i1}\fmf{electron,label=$e^{+}$}{i2,v8}
\fmf{electron,label=$e^{-}$}{o1,v7}\fmf{electron,label=$e^{+}$}{v7,o2}
\fmf{photon}{v7,v9}\fmf{photon}{v8,v10}\fmf{plain}{v9,v10}
\fmffreeze\fmfv{decor.shape=circle,decor.filled=shaded,decor.size=12}{v9,v10,v7,v8}\fmf{quark,label=$q$,l.s=right}{v9,v1} \fmf{quark,label=$q$}{v2,v10}\fmfv{decor.shape=cross,decor.size=8,label=$x$,l.a=90,l.d=.05w}{v1}\fmfv{decor.shape=cross,decor.size=8,label=$0$,l.a=90,l.d=.05w}{v2}
\end{fmfgraph*}}\nonumber\\
\nonumber\\
&&\parbox{60mm}{\begin{fmfgraph*}(60,30)
\fmfstraight\fmftop{v5,v4,v2,v1,v3,v6}
\fmfleft{i1,i2}\fmfright{o1,o2}
\fmf{electron,label=$e^{-}$}{v8,i1}\fmf{electron,label=$e^{+}$}{i2,v8}
\fmf{electron,label=$e^{-}$}{o1,v7}\fmf{electron,label=$e^{+}$}{v7,o2}
\fmf{photon}{v7,v9}\fmf{photon}{v8,v10}\fmf{plain}{v9,v10}
\fmffreeze\fmfv{decor.shape=circle,decor.filled=shaded,decor.size=12}{v7,v8}\fmfv{decor.shape=circle,decor.filled=shaded,decor.size=12,label=$q$,l.a=100,l.d=.1w}{v9}\fmfv{decor.shape=circle,decor.filled=shaded,decor.size=12,label=$q$,l.a=80,l.d=.1w}{v10}\fmf{quark}{v10,v1}\fmf{quark}{v2,v9}\fmfv{decor.shape=cross,decor.size=8,label=$x$,l.a=90,l.d=.05w}{v1}\fmfv{decor.shape=cross,decor.size=12,label=$0$,l.a=90,l.d=.05w}{v2}
\end{fmfgraph*}}\nonumber
\end{eqnarray}
\caption{Lowest order contributions.}
\end{figure}
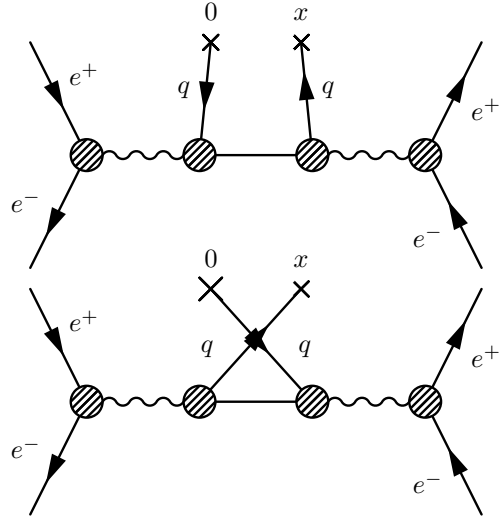
The distinction between both contributions is given by the fact, that the upper diagram has got an imaginary part, whereas the lower one is completely real so that we get only a contribution from the upper diagram 
\begin{eqnarray}
f_{inc}(p_{3})&=&\frac{1}{2(2\pi)^3w}\pi e^4Q_{q}^2 l^{\mu\nu}\left\vert\frac{1}{s-\lambda^2+i\epsilon}\right\vert^2\nonumber\\
&&2\sum_{s_{q}}\bar{u}(p_{3})\gamma_{\nu}(\pslash-M)\gamma_{\mu}u(p_{3})\nonumber\\
&&\quad\delta^{(1)}(p^2-M^2)\theta(p^0)\raisebox{-1ex}{\em $\mid_{p=p_{1}+p_{2}-p_{3}}$}.\nonumber\\
\end{eqnarray}
which is, of course, the standard result, which one obtains in considering to lowest order the reaction
\begin{eqnarray}
e^{+}+e^{-}\to q+\bar{q}
\end{eqnarray}  
for two quark flavours of charge $Q_{q}$.

\subsection{Path Integral Representation}

In the last subsection ${\cal M}$ and therefore the one-quark-inclusive cross-section could be expressed in terms of quark 6-point-functions after separating the electromagnetic interaction by a perturbative calculation. As the coupling $g$ is not assumed to be small we will now derive a representation of these quark 6-point-functions suitable for non-perturbative calculations. For this we consider the Hamiltonian path integral in our effective theory obtained from the abelian gluon model.

\medskip

Any Green's function of the theory can be written as
\begin{eqnarray}
&&\langle\!\langle 0 |\ \mbox{T}q(x_{1})... \bar{q}(x_{2})...\ | 0 \rangle\!\rangle\\
&&\qquad={\tilde{\cal Z}}^{-1}\int{\cal D}(G,\Pi_{G},q,\bar{q})\ q(x_{1})...\bar{q}(x_{2})...\nonumber\\
&&\qquad\qquad\times\exp \{i\int\! d^4\!y\ (\Pi(y)\dot{\Phi}(y)-\tilde{{\cal H}}(y))\}\nonumber
\end{eqnarray}
with
\begin{eqnarray}
{\tilde{\cal Z}}&=&\int{\cal D}(G,\Pi_{G},q,\bar{q})\nonumber\\
&&\ \times\exp \{i\int\! d^4\!y\ (\Pi(y)\dot{\Phi}(y)-\tilde{{\cal H}}(y))\}. 
\end{eqnarray}
Here the classical fields $G$ and $\Pi_{G}$ and the Grassmann fields $q$ and $\bar{q}$ have to be inserted into the part of $\tilde{\cal H}(y)$ (\ref{eqn16}) which describes the abelian gluon model. Furthermore for $y^0>0$ the classical and Grassmann fields have to be time-shifted.

\medskip

As the Hamiltonian path integral is quadratic in $q,\bar{q}$, the fermionic fields can be integrated out. With the Green's function $\tilde{S}_{F}$ of the quark in the modified gluon background
\begin{eqnarray}
(i\partialslash-M_{0}-g\Gtildeslash(z_{1}))\tilde{S}_{F}(z_{1},z_{2};G)\nonumber\\=-\delta^{(4)}(z_{1}-z_{2}),
\end{eqnarray}
the quark propagator is given as
\begin{eqnarray}
\langle\!\langle\ 0\mid Tq(z_{1})\bar{q}(z_{2})\mid 0\rangle\!\rangle=\langle\!\langle\ \frac{1}{i}\tilde{S}_{F}(z_{1},z_{2};G)\rangle\!\rangle,
\end{eqnarray}
where the brackets $\langle\!\langle\ \rangle\!\rangle$ on the right hand side denote the average over all gluon fields with the measure dictated by the path-integral
\begin{eqnarray}
\langle\!\langle F(G,\Pi_{G})\rangle\!\rangle={\tilde{\cal Z}}^{-1}\int{\cal D}(G,\Pi_{G})F(G,\Pi_{G})\nonumber\\
\times\exp \{i\int\! d^4\!y\ (\Pi(y)\dot{\Phi}(y)-\tilde{{\cal H}}(y))\}
\end{eqnarray} 
Inserting the Green's function into the matrix element we get the inclusive cross section expressed explicitely in terms of a Hamiltonian path integral. This expression should be a convenient starting point for applying non-perturbative methods to an evaluation of inclusive cross sections. It should be possible, for instance, to generalise the methods of \cite{9} to the case of this effective theory here.

\section{CONCLUSIONS}

We have written the inclusive cross section as imaginary part of an amplitude ${\cal C}$ for which we have given a path integral representation. We have applied our formalism to the reaction $e^{+}+e^{-}\to q+X$ and checked that in lowest nontrivial order we get the correct result. Our hope is that our path integral representation will lead to a genuinely non-perturbative evaluation of inclusive cross sections at high energies along the lines of \cite{9}. The generalisation to QCD should be straightforward. Our methods should allow a description of inclusive production of hadrons $h$ in $e^{+}e^{-}$ annihilation at high energies
\begin{eqnarray}
e^{+}+e^{-}\to h+X
\end{eqnarray}
in terms of a genuine OPE. A similar approach should be possible for fracture functions \cite{14} for hadron-hadron scattering
\begin{eqnarray}
h_{1}+h_{2}&\to&h_{3}+X.
\end{eqnarray}

\end{fmffile}

\end{document}